# Three-dimensional Investigation of Grain Orientation Effects on Void Growth in Commercially Pure Titanium

Marina Pushkareva[1]; Jérôme Adrien[2]; Eric Maire[2]; Javier Segurado[3,4]; Javier Llorca[3,4]; Arnaud Weck[1,3,5,6*]

[1] Department of Mechanical Engineering, University of Ottawa, 150 Louis Pasteur, Ottawa, ON K1N 6N5, Canada

[2] Université de Lyon, INSA-Lyon, MATEIS CNRS UMR5510, 7 Avenue Jean Capelle, F-69621 Villeurbanne, France

[3] IMDEA Materials Institute, C/Eric Kandel 2, 28906 Getafe, Madrid, Spain

[4] Department of Materials Science, Polytechnic University of Madrid, E. T. S. de Ingenieros de Caminos, 28040, Madrid, Spain

[5] Department of Physics, University of Ottawa, 150 Louis Pasteur, Ottawa, ON K1N 6N5, Canada

[6] Centre for Research in Photonics at the University of Ottawa, 800 King Edward Ave., Ottawa, ON K1N 6N5, Canada

*corresponding author: aweck@uottawa.ca

## Abstract

The fracture process of commercially pure titanium was visualized in model materials containing artificial holes. These model materials were fabricated using a femtosecond laser coupled with a diffusion bonding technique to obtain voids in the interior of titanium samples. Changes in void dimensions during in-situ straining were recorded in three dimensions using x-ray computed tomography. Void growth obtained experimentally was compared with the Rice and Tracey model which predicted well the average void growth. A large scatter in void growth data was explained by differences in grain orientation which was confirmed by crystal plasticity simulations. It was also shown that grain orientation has a stronger effect on void growth than intervoid spacing and material strength. Intervoid spacing, however, appears to control whether the intervoid ligament failure is ductile or brittle.

# Introduction

A better understanding of damage phenomena in hexagonal close-packed (HCP) metals is needed to improve the prediction capabilities of fracture models. To date, most of the research has focused on understanding the deformation mechanisms in some of these HCP materials, including titanium [1-10] for aerospace applications, zirconium [11-14] for the nuclear industry and magnesium [15-17] for manufacturing industries. The deformation behavior of HCPs is quite complex as it is affected by twinning and grain orientation. Various studies have shown that grain orientation can significantly affect deformation and mechanical properties [2, 3, 18-26]. In deciding the deformation mode in HCP materials one should consider the c/a ratio [5]: (i) when c/a < 1.632 (Ti, Zr, Be, etc), a sufficient number of slip systems is provided by prismatic and pyramidal slip; (ii) when 1.63 < c/a < 1.73 (Mg), only one basal slip system is available; and (iii) when c/a > 1.73 (Zn, Cd) deformation is accommodated by basal slip and twinning. Propensity of twinning is also related to the grain size [4]: smaller grains cannot accumulate dislocations and twin boundaries, and hence twinning decreases. Some work on fracture in HCP materials showed that ductility is also affected by the HCP crystal structure, grain orientation and twinning. Yoo [1] proposed a relation between ductility and the ratio of elastic bulk modulus to shear modulus, K/G. High value of K/G were associated with ductility (titanium, zirconium) and a low value with brittleness (magnesium). Fracture in Ti was also proposed to be a result of void nucleation, growth and coalescence. Krishna Mohan Rao et al. [10] reported fracture at room temperature by void nucleation and growth in a near-α aircraft titanium alloy. It was shown that during fatigue testing, a crack nucleated at a hard grain – soft grain interface propagates through the hard-orientated grain, suggesting that the driving stress intensity depends on the local morphology and crystallographic orientations [27]. Furthermore, the plastic zone which develops at the tip of a crack in an hcp single crystal is strongly dependent on crystallographic orientation

[27]. Even grains with similar orientations can follow different orientation trajectories depending on their misorientation with respect to neighboring grains [28]. The importance of neighboring grains can be understood in terms of the stress intensities resulting from grain misorientations [27].

A review on the effect of twinning, grain orientation and other parameters on fracture processes (experiments and modeling) was provided by Roters et al. [29]. The effect of grain orientation on the void growth process has been numerically studied at different length scales: at the nanometric scale using Molecular Dynamics [30, 31], at the microscale by discrete dislocation dynamics [32-34] and at the continuum level using crystal plasticity [35]. In all these studies the general conclusion is that under uniaxial loading, void growth in single crystals slightly depends on the orientation of the tensile axis with respect to the crystalline lattice.

Most experimental studies in literature are devoted to FCC materials, and orientation effects in HCP metals are lacking. Models have been proposed to predict deformation mechanisms, evolution of crystallographic structure, texture, twinning, hardening and mechanical properties during deformation [6-9]. However there is a lack of models to predict fracture in HCP structures.

From the results presented in the literature, fracture in Ti can be the result of a ductile fracture process where voids are nucleated, grow and finally coalesce. The approach taken in this paper is to verify whether standard ductile fracture models, which were proven successful for some isotropic materials [36], are able to predict fracture and in particular void growth in an HCP structure. Experimental data on void growth and coalescence are difficult to obtain, making model validation difficult. To simplify the study of ductile fracture, laser-drilled model materials similar to those of Weck et al. [37] were produced out of commercially pure titanium, tested, and visualized using x-ray tomography. Void growth was followed in-situ and compared to the Rice and Tracey model for void growth, and to crystal plasticity simulations.

# 1 Experimental methods

Commercially-pure titanium (Grade 1) was purchased from the company NewMet in the form of thick 0.25 mm titanium foil (99.8% Ti) and AlfaAesar in the form of thin 0.032 mm titanium foil (99.7% Ti). Arrays of voids were introduced in the thinner titanium foils using femtosecond laser micromachining, a technique which allows precise manufacturing of voids without the formation of a heat affected zone. This technique has already been successfully implemented to create model material out of copper and aluminum [37-39]. Different void configurations were used: (i) rectangular array with intervoid spacings of 70 µm, 100 µm and 120 µm, and (ii) holes at 45° with intervoid spacing of 70 µm, 100 µm and 130 µm. All samples had an initial void diameter of 35 µm. After laser drilling, the samples were polished down to a mirror finish using a 0.05 µm colloidal silica suspension. Sheets containing voids were then diffusion bonded to sheets without voids in a vacuum furnace at $10^{-5}$ torr. Different annealing temperatures and times were used in order to obtain different levels of strength, from 950 °C for 2h to 1000 °C for 3h. Two types of samples were obtained: one with a 2-dimensional array of voids inside the material and one with a 3-dimensional array of voids to obtain a more realistic void configuration. Sample dimensions are shown in Figure 1.

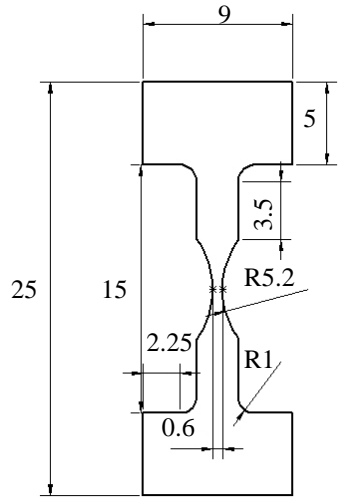

Figure 1. Sample dimensions in mm. Sample thickness was approximately 0.5 mm.

Titanium samples were pulled in-situ in an x-ray tomography system available at Mateis lab, INSA de Lyon, France, and described in [40]. X-ray tomographs were acquired on a CCD camera with 960 x 768 sensitive elements with a 1.5 µm voxel size. An x-ray energy of 80 keV was used. The number of projections used was 720 initially and was then decreased to 180 in order to decrease time of acquisition while maintaining acceptable image quality. The acquisition time was 500 ms per rotation angle with a total rotation of 180°. Tensile tests were performed at room temperature on a specially designed in situ tensile testing machine. The testing speed was 1 µm/s at the beginning of the test and was decreased to 0.5 µm/s close to the end of the test in order to capture void linkage. The machine was used in tension with a load cell of 2 kN and both load and displacement were recorded during the test. During the in-situ tomography experiments, the samples were deformed in tension and the test was stopped at various levels of deformation in order to acquire tomograms. Depending on how quickly the sample failed and on the number of interesting events occurring during the test, the number of tomograms acquired varied between samples (5 to 10 tomograms).

# 2 Experimental results and discussion

### 2.1 Microstructural examination

Metallographic examination of all samples reveals fine features of platelike α (Figure 2). As the samples were all diffusion bonded at temperatures higher than the α to β phase transformation temperature, plate-like alpha is expected because the cooling rate at the phase transformation temperature (882 °C) was about 0.40-0.46 °C/s [41]. The width of the α plates was found to slightly increase with decreasing diffusion bonding temperature and time (Table 1, Figure 2). It is important to note here that the microstructure consists of colonies of α-plates (or α-lamellas) which form grains. This means that inside each grain, α-plates have the same orientation [42]. Thus the microstructural unit of deformation is the grain (which is larger than the void) whose dimensions are reported in Table 1.

Values of microhardness measured on samples annealed at different temperatures are presented in Table 1. Longer annealing results in slightly smaller α width and in significantly higher values of microhardness. While the α plates width could contribute to the increase in strength, the small change in plate width cannot explain the large change in hardness observed experimentally. The increase in hardness is therefore deduced to be a result of oxidation of the titanium samples during heat treatment. Higher annealing temperature and time result in higher oxygen content in titanium thus higher microhardness, as already demonstrated in [43]. Indeed, oxygen addition leads to solid solution strengthening by asymmetric lattice distortion [44]. The range of oxygen calculated based on the data provided in [43] was found to be ranging from 0.035 at % for sample S2 to 0.044 at % for sample S6.

Table 1. Width of platelike α, α colony size, grain size and microhardness of samples after diffusion bonding.

| Diffusion bonding parameters | | α width, μm | Δα width, μm | α colony size *grain size, μm | Δα colony size *Δgrain size, μm | HV 0.2 | ΔHV 0.2 |
| --- | --- | --- | --- | --- | --- | --- | --- |
| Temperature, °C | Time, hours | | | | | | |
| 1000 | 3 | 16 | 3 | 104 | 10 | 252 | 32 |
| 1000 | 2 | 17 | 4 | 92 | 10 | 238 | 28 |
| 960 | 2 | 18 | 4 | 86 | 10 | 228 | 17 |
| as-received | | N/A | N/A | *10 | *4 | 178 | 11 |

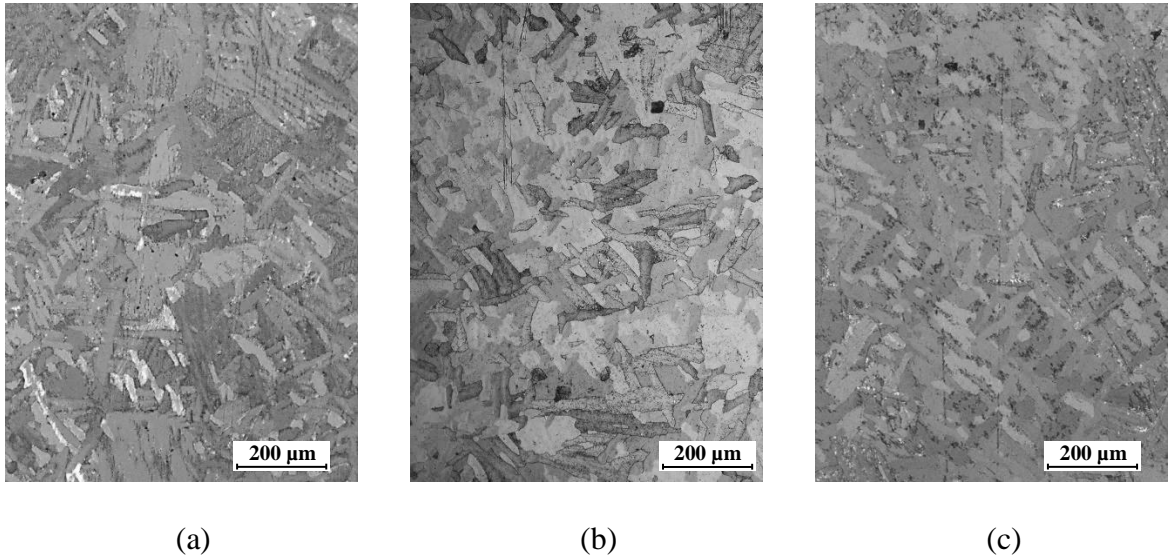

(a)          (b)          (c)

Figure 2. Typical sequence of microstructures (×100) versus diffusion bonding temperature and time: (a) 960 °C, 2 hours, (b) 1000 °C, 2 hours, (c) 1000 °C, 3 hours.

### 2.2 In-situ Tensile Test Results

Using the force registered during the tensile test and the smallest cross sectional area extracted from tomographic reconstructions, true stress-strain curves were constructed for all samples. The smallest cross sectional area was calculated using ImageJ [45] by filling the holes and taking the slice normal to the tensile axis with the smallest area. True stress-strain curves were then calculated using the following equations:

$$\sigma = \frac{L}{A} \qquad (1)$$

$$\varepsilon = \ln\left(\frac{A_0}{A}\right) \qquad (2)$$

where $L$ is the load; $A$ is the current cross sectional area; $A_0$ is the initial cross sectional area. True stress-strain curves are represented in Figure 4. All true stress-strain curves obtained were fitted using the classical Hollomon expression:

$$\sigma = K\varepsilon^n \qquad (3)$$

where $K$ is a constant and $n$ is the hardening exponent. The macroscopic necking strain is given by the hardening exponent $n$. Hardening exponents ranging from 0.09 to 0.16 were extracted from the stress strain curves (Table 2). The schematic representation of two void configurations is shown in Figure 3.

One can note (Figure 4) that the yield stress of samples annealed at 960 °C is lower than that of samples annealed at 1000 °C. This correlates with the microhardness results presented in section 2.1 which were explained in terms of sample oxidation during annealing. Also the hardening exponent of samples annealed at 960 °C is higher than that of samples annealed at 1000 °C (Table 2).

Table 2. Hardening exponent *n* of different samples.

| Sample | Diffusion bonding parameters | | Void configuration | | Number of layers | $n$ |
|---|---|---|---|---|---|---|
| | Temperature, [°C] | Time, [hours] | Void interspacing, [μm] | Angle, [°] | | |
| S1 | 960 | 2 | 99 | 45 | 1 | 0.15 |
| S2 | 960 | 2 | 94 | 45 | 3 | 0.16 |
| S3 | 1000 | 2 | 96 | 45 | 1 | 0.11 |
| S4 | 1000 | 2 | 133 | 45 | 1 | 0.14 |
| S5 | 1000 | 3 | 65 | 0 | 1 | 0.09 |
| S6 | 1000 | 3 | 120 | 0 | 1 | 0.10 |

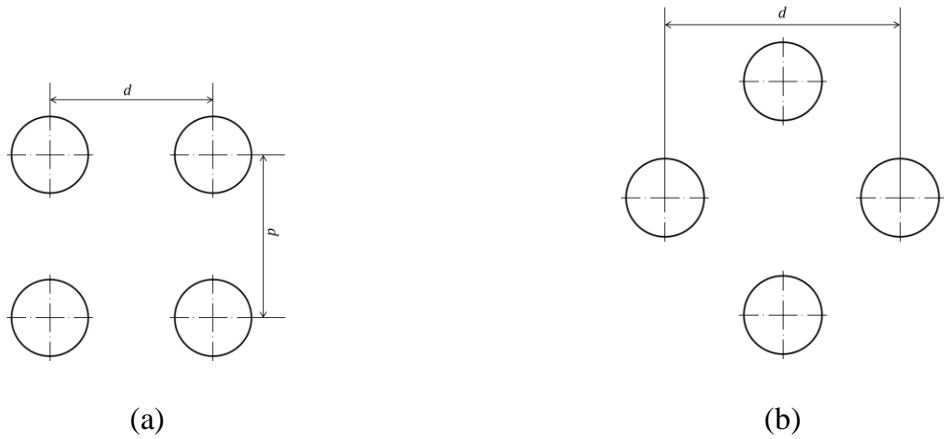

(a)  (b)

Figure 3. Schematic representation of two void configurations: (a) rectangular array with intervoid spacing d; (b) voids at 45° with intervoid spacing d.

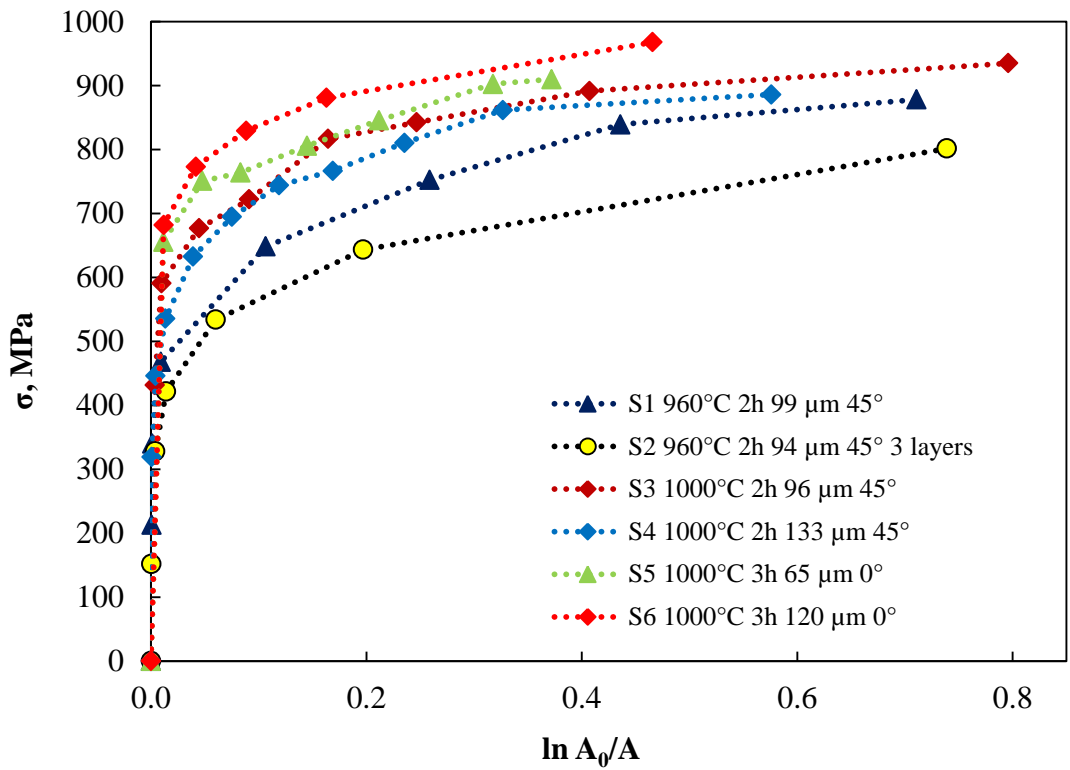

Figure 4. True stress-strain curves of CP titanium samples tested in tomography. In the legend, the name of the sample is followed by diffusion bonding temperature and time, and by the void configuration parameters (intervoid spacing and angle of the voids with respect to the tensile axis).

2.3 X-ray tomograms

X-ray tomography allowed the visualization of void growth and coalescence in three dimensions. To visualize voids in three dimensions, the open source image analysis software ImageJ was used [45]. Void dimensions as a function of deformation were extracted manually. A typical example of tomograms is shown in Figure 5 (case of the S5 sample) where the evolution of void dimensions can be clearly observed. Voids first elongate in the tensile direction during the test and elongate in the transverse direction when coalescence starts. A sequence of tomographic reconstructions is shown in Figure 6 for sample S5. Void growth was quantified in the different samples as a function of true strain (Figure 7) and the following observations can be made: (i) when comparing samples S1 and S3, where the main difference between the two samples is the annealing temperature (i.e. material strength), it can be seen that void growth is similar in the harder material (1000 °C 2h) and the softer material (960 °C 2h) because the void growth curves overlap each other. This result indicates that the material strength does not significantly affect void growth or that another effect counterbalances its influence such as the grain orientation. (ii) Increasing the intervoid spacing for voids at both 45° and 0° leads to faster void growth (see samples S3 and S4, and S5 and S6 respectively) which seems counterintuitive. Again, this suggests that void growth may be more affected by the grain orientation in which voids are located rather than by the intervoid spacing. This effect is more evident in Figure 8 where the scatter in void growth for a given sample is large and is probably due to grain orientation effects.

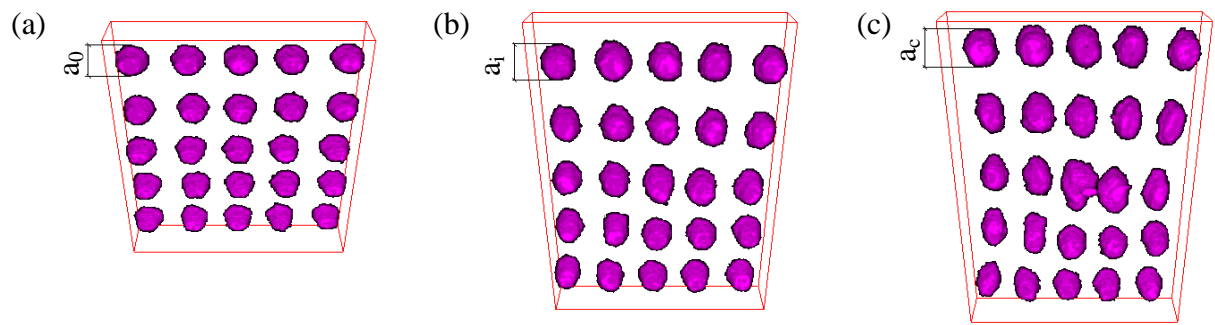

Figure 5. Tomographic reconstruction of a titanium sample containing one array of laser drilled holes. The example is for the S5 sample, that has been tested in-situ and the tomograms correspond to the following true strains (a) 0.00, (b) 0.16, (c) 0.30. The tensile direction is vertical. Initial void diameter $a_0$ is 35 μm.

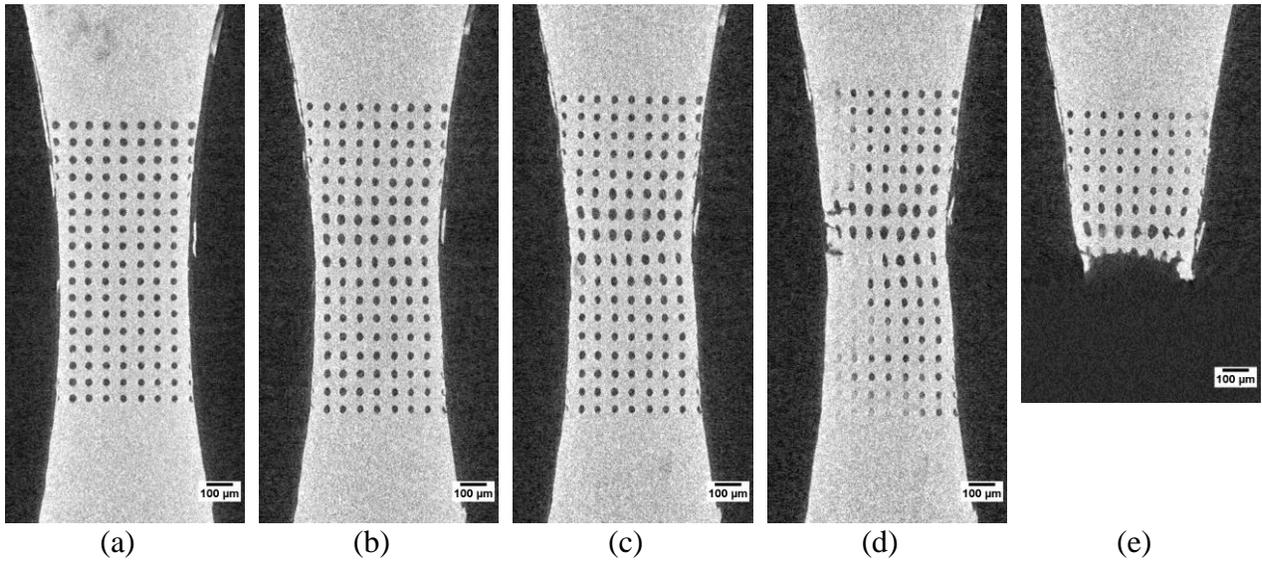

(a) (b) (c) (d) (e)

Figure 6. Sequence of tomographic reconstructions showing deformation for a titanium sample (S5) with a rectangular array of voids oriented at 0° with respect to the tensile axis at the following true strains: (a) 0, (b) 0.15, (c) 0.21, (d) 0.32 and (e) at fracture. Tensile axis is vertical.

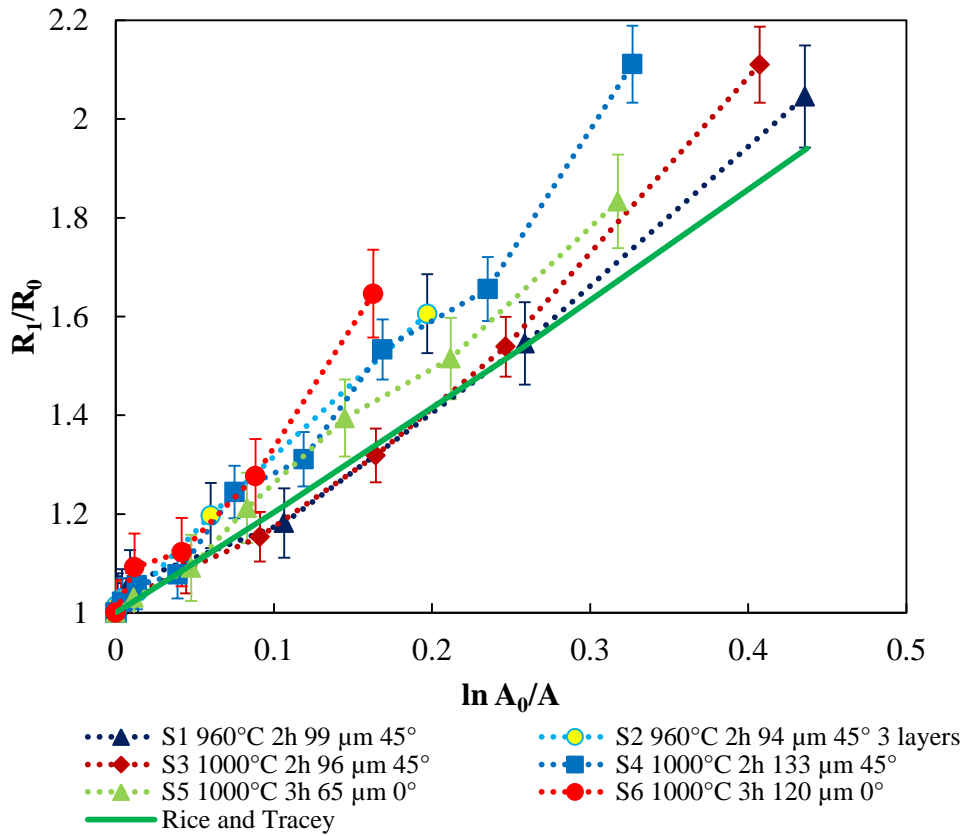

Figure 7. Void growth in the tensile direction for the fastest growing void.

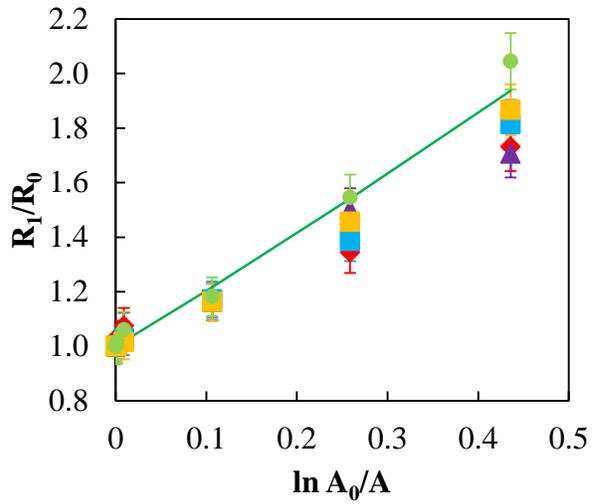
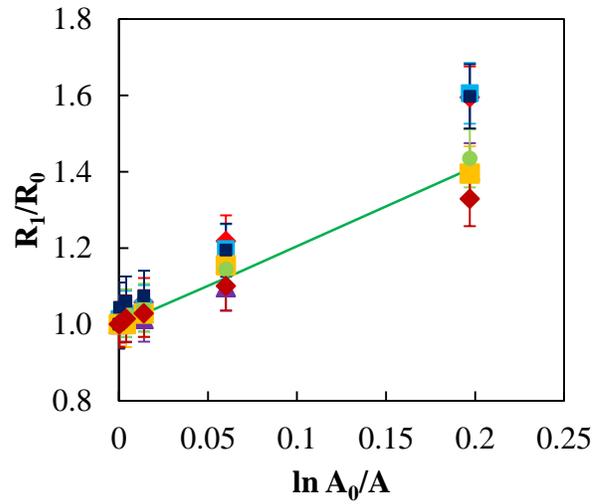
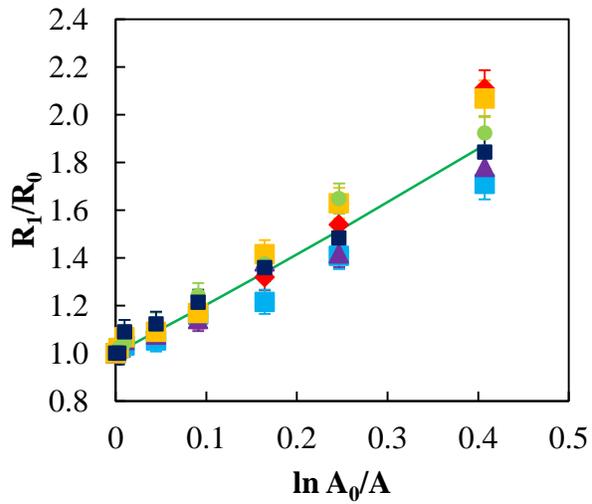
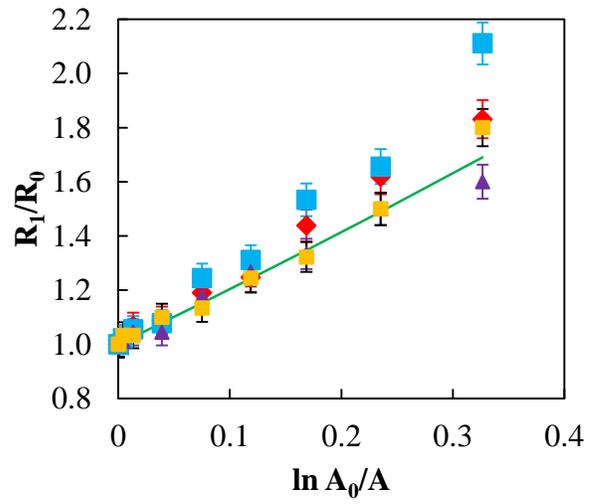

(a)

(b)

(c)

(d)

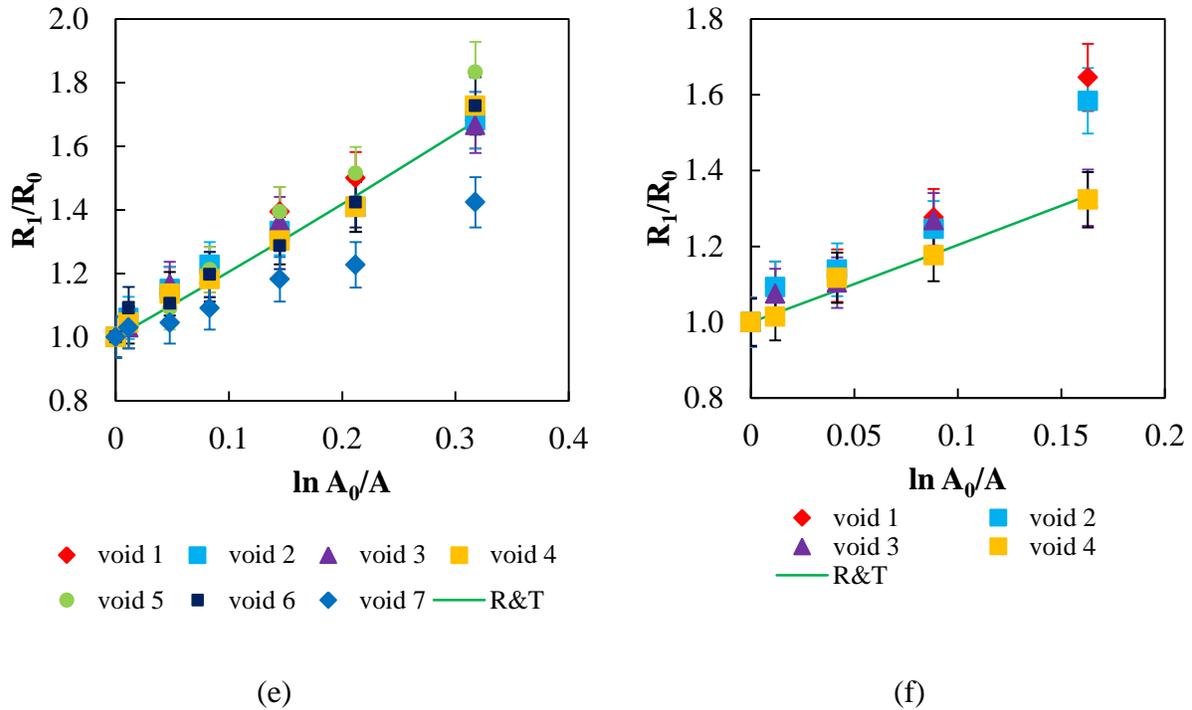

(e)            (f)

Figure 8. Comparison between experimental void growth (for voids participating in the fracture surface) and the Rice and Tracey model for samples (a) S1, (b) S2, (c) S3, (d) S4, (e) S5, (f) S6.

2.4 Fractography

Fracture surfaces of titanium samples tested in tomography were observed in a scanning electron microscope (SEM) and are shown in Figure 9. Figure 9(a) shows that the failure of the ligament between voids in samples S4 is largely ductile with the ligament necking down to a line. In samples S1 (Figure 9(b)) one can see that the ligament between voids appears less uniform and contains more secondary voids. Sample S5 (Figure 9(c)) reveals an intervoid ligament failure with a brittle character as cleavage facets can be observed between most of the voids. One important difference between samples S4, S1 and S5 is their intervoid spacing. Based on the fracture analysis described above, a larger intervoid spacing (sample S4) results in a more ductile ligament failure while smaller intervoid spacing (sample S5) leads to a more brittle mode of ligament failure. When voids are close to each other there is a high probability for having only one grain in the intervoid ligament which would favor uninterrupted strain localization between voids and thus cleavage along a given slip system. If voids are further apart, more grains will be present in the intervoid ligament and the continuity of strain localization will be broken, making it more difficult for the grains to cleave. Furthermore, the closer the voids are, the less constraint the intervoid ligament will be, up to the point where localization bands can form freely from one void to the other if the voids are close enough (similar to reaching a Brown and Embury type criterion [46]). More work would need to be done in this direction to better understand the role on fracture of the intervoid spacing and number of grains in the ligament between voids.

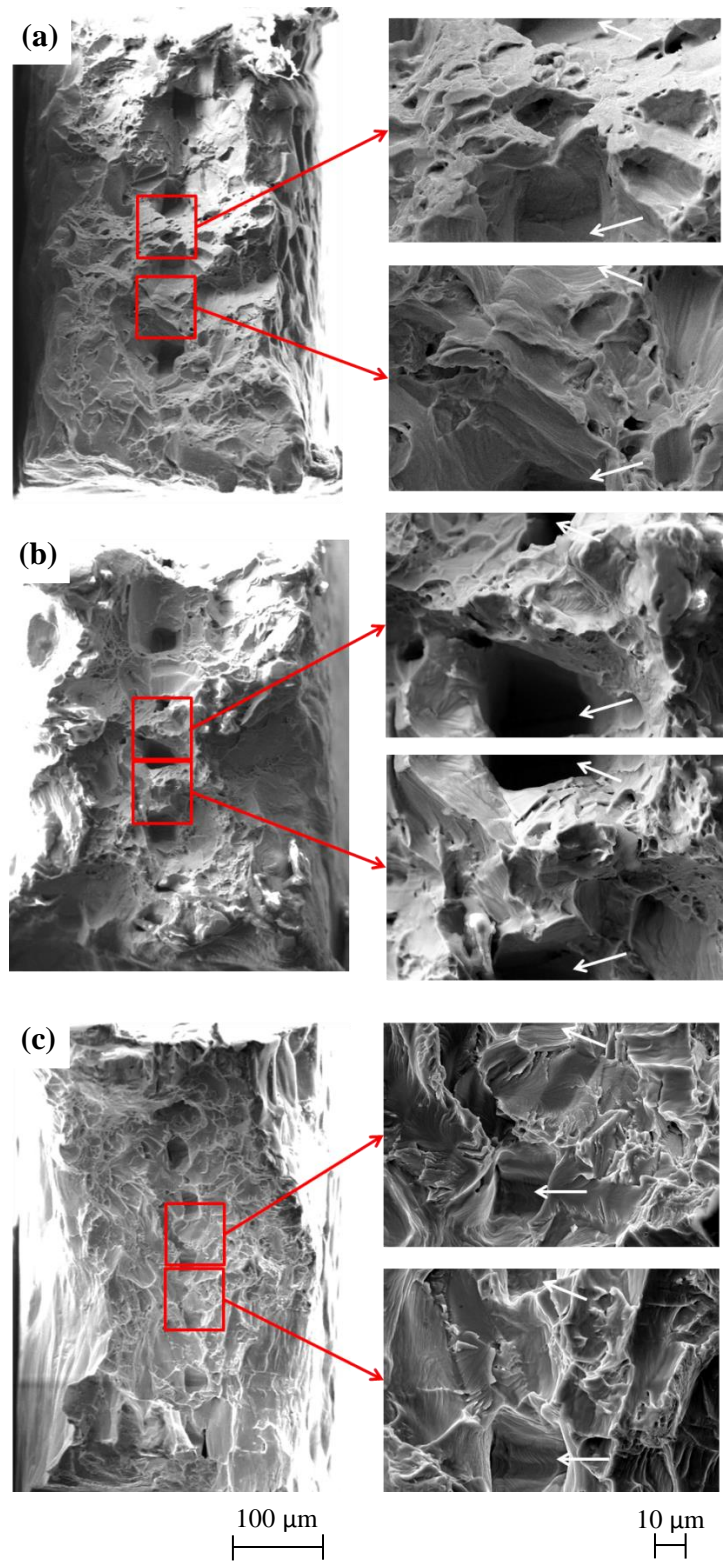

Figure 9. SEM images of the fracture surface of (a) sample S4 (b) sample S1, and (c) sample S5 with their respective close-ups of the ligament between two laser voids (indicated by the white arrows).

# 3 Comparison with existing models

Experimental results were compared to the Rice and Tracey model [47] for void growth. The Rice and Tracey equations give the principal radii $R_1$ of the ellipsoidal void in the case of uniaxial tension in the tensile direction [36]:

$$R_1 = \left[exp(D\varepsilon_1) + \frac{1+E}{D}exp(D\varepsilon_1) - 1\right]R_0 \qquad (4)$$

where

$$D = 0.558 sinh\left(\frac{3}{2}\frac{\sigma_m}{Y}\right) + 0.008 cosh\left(\frac{3}{2}\frac{\sigma_m}{Y}\right) \qquad (5)$$

$\varepsilon_1$ is the total logarithmic strain integrated over the total strain path, $\sigma_m$ is the mean stress and $Y$ the yield stress. The parameters $D$ and $E$ are used in the Rice and Tracey model to vary the contribution of the volume and shape changing part to void growth respectively. In the case of pure titanium, because the difference between the yield stress and the ultimate tensile stress (UTS) is high (more than a factor of 2), the material is assumed to be strongly hardening and the parameter $(1 + E)$ from equation (4) is equal to 5/3 as proposed by Rice and Tracey [47].

The radius of curvature and the radius of the minimal section were measured in order to determine the average stress triaxiality $T$ in the center of the minimum cross-section, using the Bridgman formula [48]:

$$\frac{\sigma_m}{Y} = \frac{1}{3} + \ln\left(\frac{a+2r}{2r}\right) \qquad (7)$$

where $a$ is the radius of the smallest cross sectional area, and $r$ the neck radius.

Stress triaxiality does not evolve significantly in the samples and is equal to 0.37±0.003.

When compared to the fastest growing void leading to fracture in all samples (Figure 7), the Rice and Tracey model appears to provide a lower bound for void growth. When compared to all voids in a given sample, the Rice and Tracey model provides good overall void growth predictions (Figure 8). However, there is a large scatter in the experimental void growth results which means that different voids grow at different rates, possibly due to grain orientation effects. Similar scatter was observed in Ti-6Al-4V alloy [49]. To better understand and quantify grain orientation effects on void growth, crystal plasticity simulations were carried out.

## 4 Crystal plasticity simulations

The effect of grain orientation on void growth in titanium has been studied using a crystal plasticity finite element model. The behavior of Ti crystal is taken into account using an elasto-viscoplastic phenomenological crystal plasticity model which includes the microscopic mechanisms of plastic deformation by slip along basal, prismatic and pyramidal a + c systems [50]. The rate-sensitivity exponent was the same for all slip systems and was selected as m = 0.1, value obtained from [50]. The hardening is accounted for by a Voce-hardening law [51]:

$$h(\Gamma) = h_0 + \left(h_0 - h_s + \frac{h_0 h_s \Gamma}{\tau_s}\right) \exp^{-h_0 \Gamma / \tau_s}$$

where $h(\Gamma)$ is the hardening modulus, $\tau_0$ is the initial critical resolved shear stress, $\tau_s$ is the saturation shear stress, $h_0$ is the initial hardening moduli, and $h_s$ is the saturation hardening moduli.

These parameters were obtained combining literature values and inverse analysis of the experimental macroscopic tensile response (Figure 10). First, the following ratios of the CRSS between different slip systems were taken from microtests performed on crystalline pure Ti grains [52], $\tau^{basal} = 1.155 \tau^{prismatic}$ and $\tau^{pyramidal} = 2.618 \tau^{prismatic}$. Second, the ratio between $\tau_0$ and $\tau_s$ were kept constant for all slip systems. Third, the hardening moduli $h_0$ and $h_s$ were identical for the three active slip systems. Finally, with these restrictions, the three parameters remaining ($\tau_0$, $h_0$ and $h_s$) are searched imposing that the response of a polycrystalline crystal plasticity finite element model including the actual texture coincides with the corresponding experimental tensile curve (Figure 10). The experimental tensile curves of the three materials considered (samples S1, S3 and S6) are used to obtain the values of $\tau_0$, $h_0$ and $h_s$ for each material and the resulting parameters are shown in Table 3.

Table 3: Parameters of Voce-hardening law for samples S1, S3 and S6.

| Slip system | $\tau_0$ (MPa) | | | $\tau_s$ (MPa) | | | $h_0$ (MPa) | | | $h_s$ (MPa) | | |
|---|---|---|---|---|---|---|---|---|---|---|---|---|
| | S1 | S3 | S6 | S1 | S3 | S6 | S1 | S3 | S6 | S1 | S3 | S6 |
| Prismatic | 123 | 147 | 212 | 274 | 266 | 235 | 310 | 500 | 300 | 1 | 5 | 54 |
| Basal | 143 | 169 | 245 | 317 | 308 | 272 | 310 | 500 | 300 | 1 | 5 | 54 |
| Pyramidal a+c | 323 | 384 | 555 | 718 | 698 | 616 | 310 | 500 | 300 | 1 | 5 | 54 |

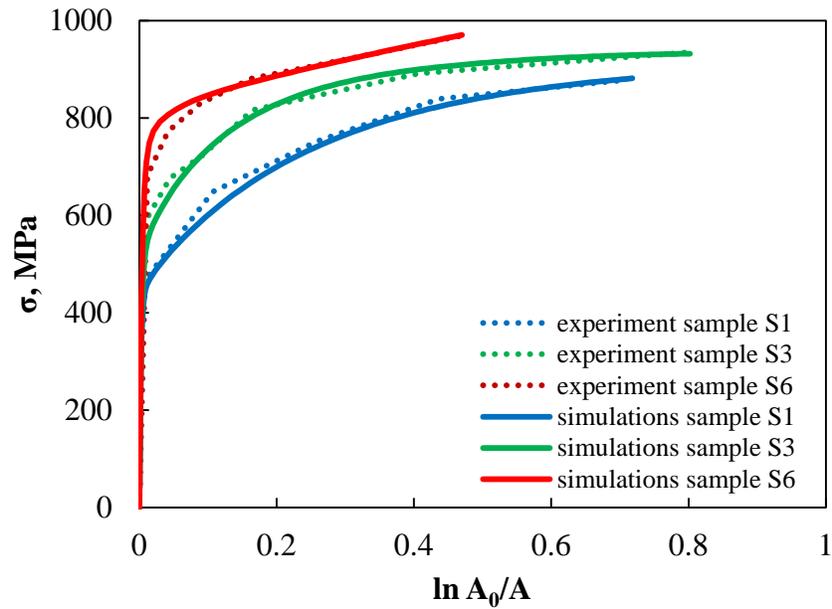

Figure 10. Tensile stress-strain curves: experimental and numerical simulations results.

Void growth simulations were run using a 3D finite element model of a grain containing a spherical void (Figure 11(a)). The grains are simulated in the shape of a tetrakaidecahedron (Figure 11(b)) which is a reasonable approximation of grains in polycrystals [53]. The grain in the middle, containing the spherical void, was modeled using the CP model previously described and using three different orientations of the crystal lattice with respect to the loading direction. The surrounding grains were modeled using standard isotropic $J_2$ plasticity with yield stress and isotropic hardening evolution taken from tensile experiments (Figure 10).

The results of the simulations show first that the evolution of void size with applied strain depends on the orientation of the grain in which the void is embedded (Figure 12, Figure 13). While the simulation results do not match exactly the experiments, the range of scatter in the results is similar in both cases indicating that the scatter observed experimentally could indeed be due to grain orientation effects. Simulated void growth results for samples S1, S3, and S6 which were annealed at different temperatures, and thus have different strength levels, are compared in Figure 14 (a). The results show that material property (in this case the strength of the material) has a much smaller effect on voids growth than the orientation of the grains in these materials, which support the experimental observations.

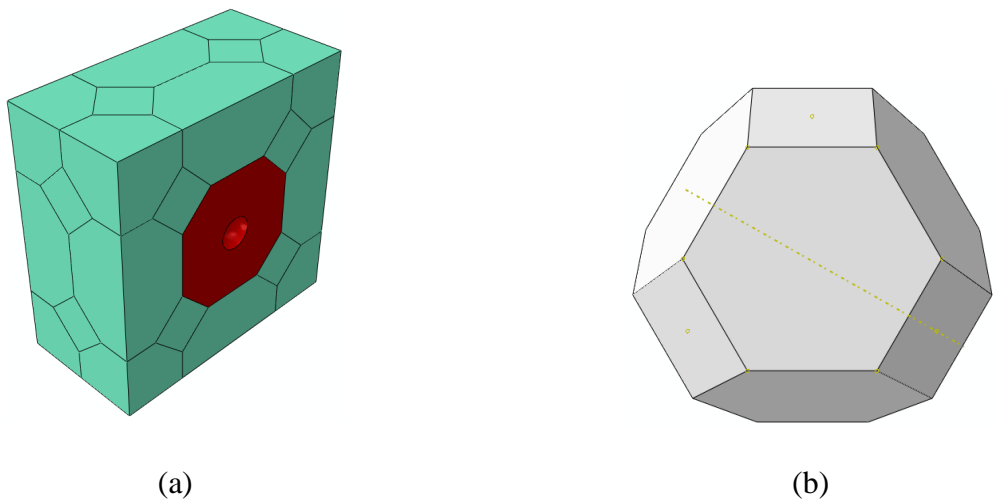

(a)           (b)

Figure 11. (a) 3D model. Half of the model is shown. Red color indicates crystal plasticity, green color indicates macroscopic properties of titanium; (b) tetrakaidecahedron used to construct the 3D model.

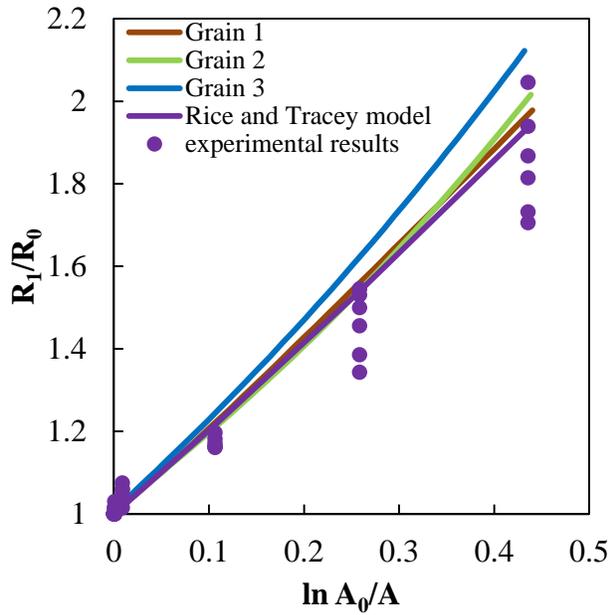

(a)

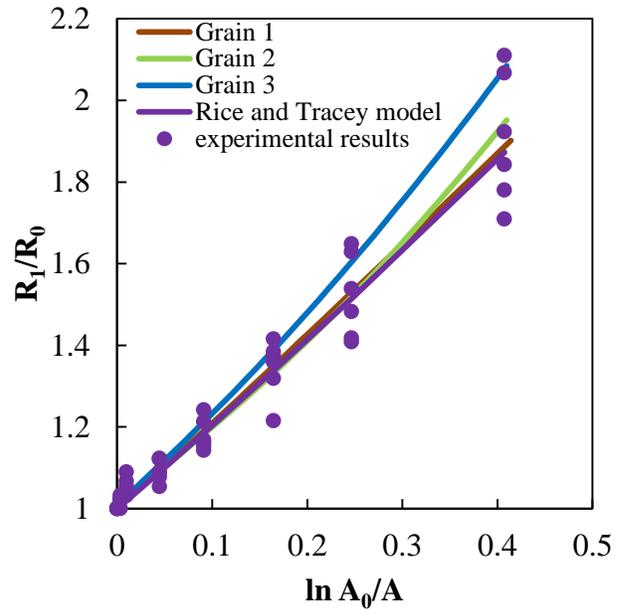

(b)

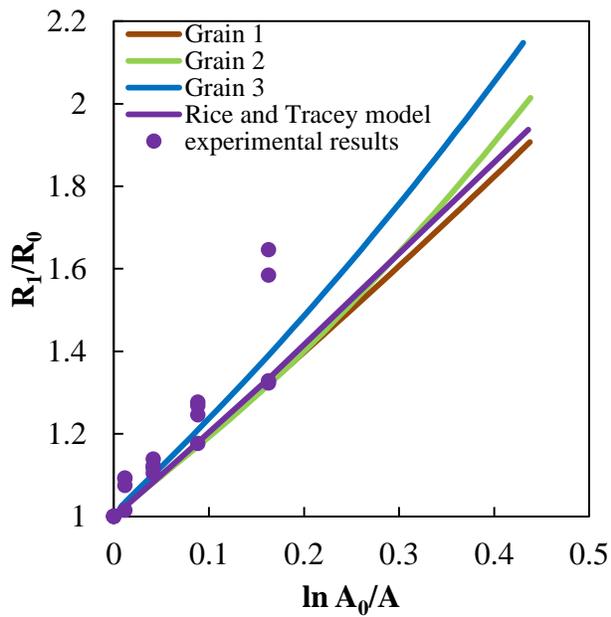

(c)

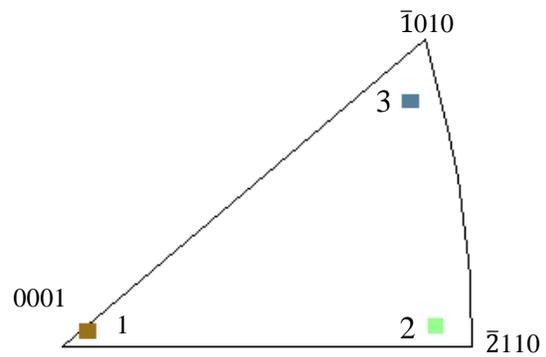

(d)

Figure 12. Experimental and simulation results of void growth in three samples: (a) S1, (b) S3 and (c) S6. (d) The inverse pole figure shows the three different grain orientations used in the simulations.

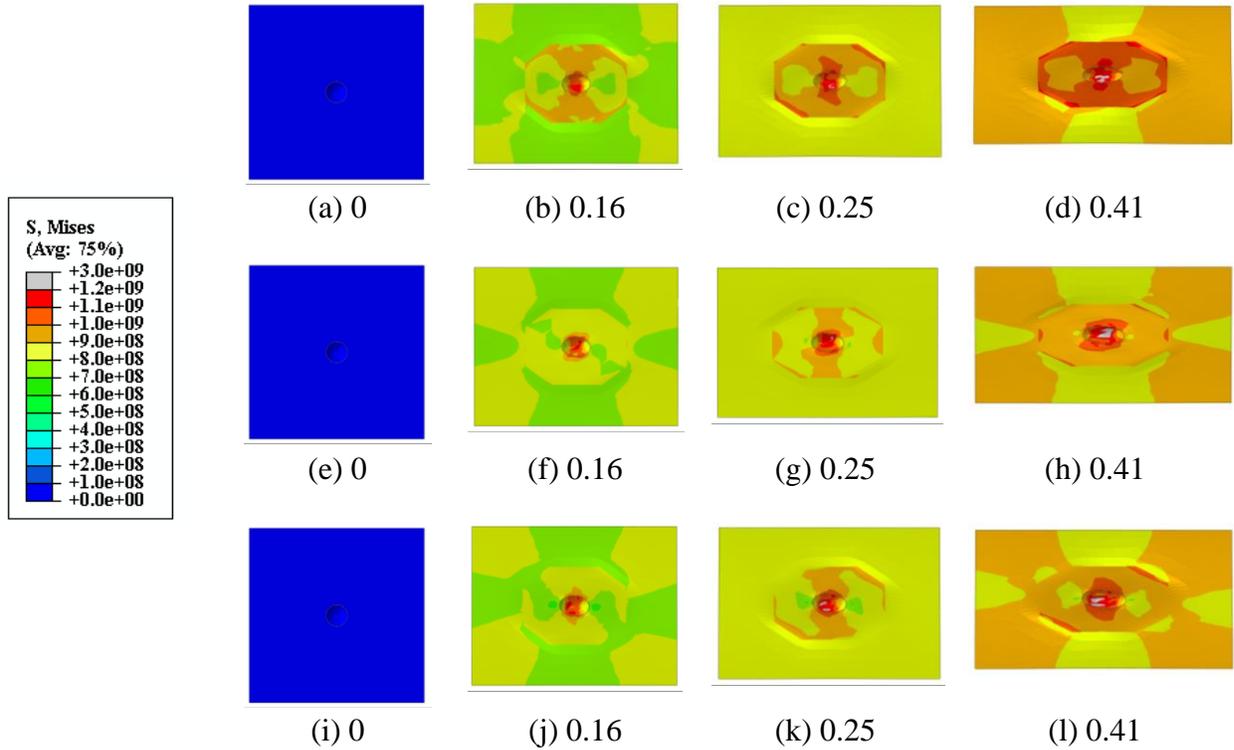

Figure 13. Visualization of void growth for sample S3 in grains with different orientation at different true strains: (a)-(d) Grain 1, (e)-(h) Grain 2, (i)-(l) Grain 3. Tensile direction is horizontal.

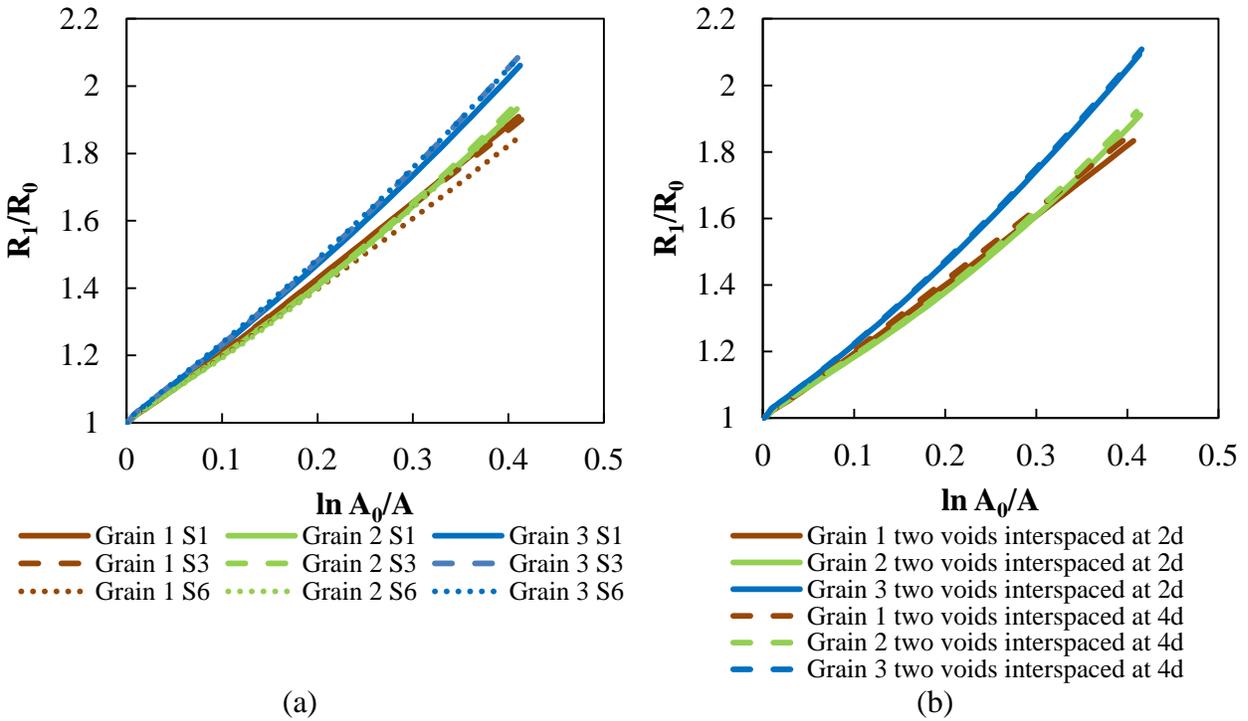

Figure 14. (a) Comparison of void growth simulation results between different samples S1, S3 and S6. (b) Comparison of void growth simulation results for sample S3 for two voids at different interspacings.

In order to understand the effect of void interspacing on void growth, crystal plasticity simulations were performed using the material properties of sample S3 for two different intervoid spacings: 2d and 4d, where d is the void diameter. To achieve this, models with two voids spaced 2d and 4d were created. The results (Figure 14 (b)) show that increasing the intervoid spacing leads to a slight increase in void growth which is in agreement with the experimental results (Section 2.3) which initially seemed counterintuitive. However, the orientation of the grain has a much stronger effect on void growth than the intervoid spacing. Note that some void rotation with respect to the tensile axis can be observed in the simulation results (Figure 13-Figure 15-Figure 16). Some void rotation can also be observed in the experimental results in Figure 6 but no direct comparison can be made with the simulation results as the grain orientation is not known in the experiments.

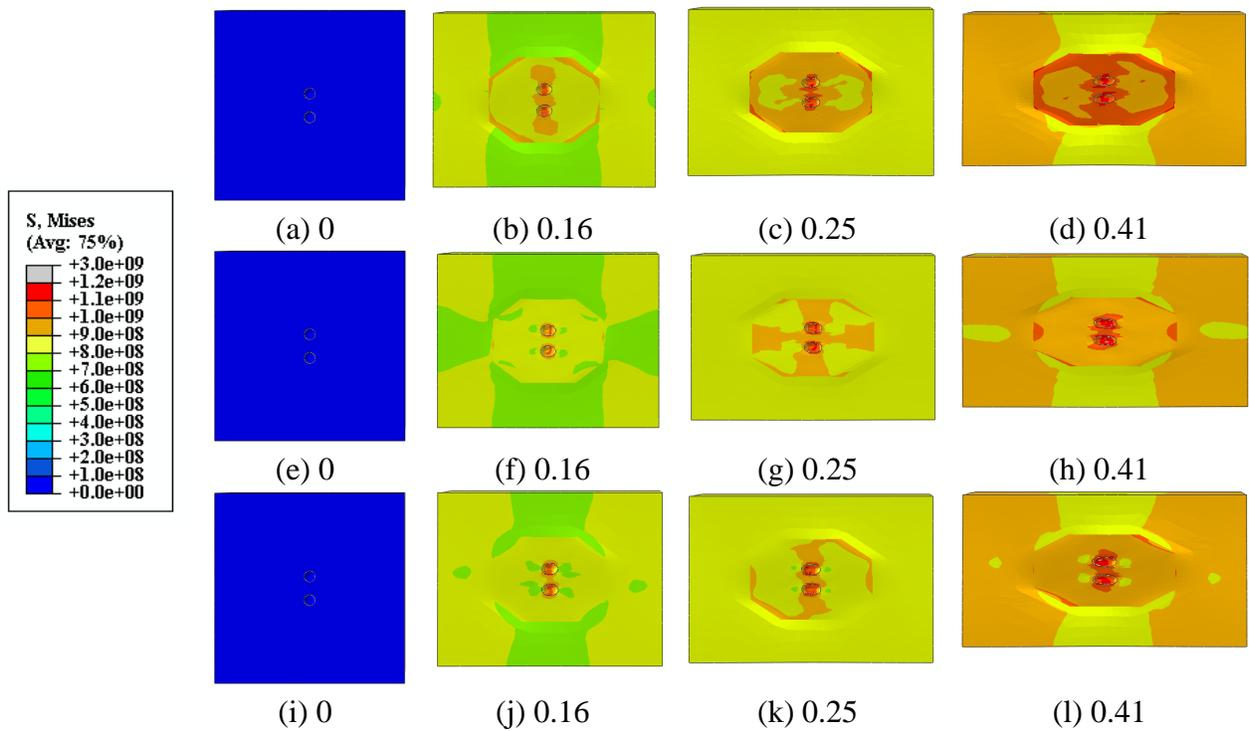

Figure 15. Visualization of void growth for two voids interspaced at 2d for sample S3 in grains with different orientation at different true strains: (a)-(d) Grain 1, (e)-(h) Grain 2, (i)-(l) Grain 3. Tensile direction is horizontal.

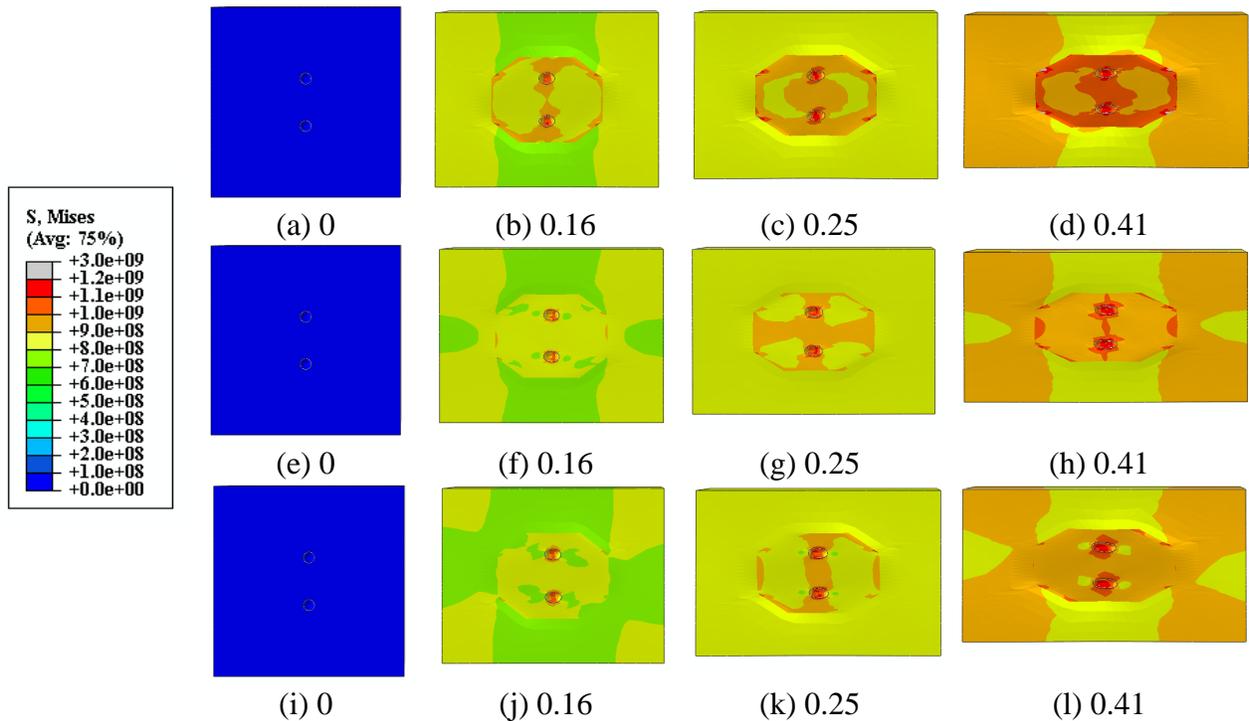

Figure 16. Visualization of void growth for two voids interspaced at 4d for sample S3 in grains with different orientation at different true strains: (a)-(d) Grain 1, (e)-(h) Grain 2, (i)-(l) Grain 3. Tensile direction is horizontal.

## Conclusion

1. Using laser-drilled materials and x-ray computed tomography, void growth was captured in detail inside titanium samples.

2. Intervoid spacing and material strength were found not to affect void growth significantly.

3. The Rice and Tracey model for void growth provided good predictions of average void growth rates.

4. Fracture surface analyses revealed different modes of intervoid ligament failure where smaller intervoid spacing leads to a more brittle failure. This effect was related to the number of grains in the intervoid ligament and to the ability of strain localization bands to form between voids.

5. A large scatter in void growth was observed experimentally and was hypothesized to be due to grain orientation effects. This hypothesis was supported by crystal plasticity finite element simulations which showed a similar scatter in void growth results.

## Acknowledgements

AW acknowledges support from the Natural Sciences and Engineering Research Council of Canada (NSERC). The International Fellowship MICROFRAC (Visualization and modeling of fracture at the microscale) supported by the European Union, H2020 program, Marie Skłodowska-Curie Actions (contract 659575) is also acknowledged. JL acknowledges support from the European Research Council (ERC) under the European Union's Horizon 2020 research and innovation program (Advanced Grant VIRMETAL, grant agreement No. 669141).